\documentclass[10pt,letterpaper]{article}

\usepackage{mascots04,epsfig,amsmath,amssymb,url,bibspacing,tweaklist}

\renewcommand{\itemhook}{\setlength{\topsep}{0pt}
\setlength{\itemsep}{0pt}}

\begin{document}

\title{
    Toward Compact Interdomain Routing
}

\author{
    Dmitri Krioukov \and kc claffy \\
    CAIDA \\
    \{dima,kc\}@caida.org
}

\maketitle

\begin{abstract}

Despite prevailing concerns that the current Internet interdomain
routing system will not scale to meet the needs of the 21st century
global Internet, networking research has not yet led to the
construction of a new routing architecture with satisfactory and
mathematically provable scalability characteristics. Worse,
continuing empirical trends of the existing routing and topology
structure of the Internet are alarming: the foundational principles
of the current routing and addressing architecture are an inherently
bad match for the naturally evolving structure of Internet
interdomain topology. We are fortunate that a sister discipline,
theory of distributed computation, has developed routing algorithms
that offer promising potential for genuinely scalable routing on
realistic Internet-like topologies. Indeed, there are many recent
breakthroughs in the area of {\em compact routing}, which has been
shown to drastically outperform, in terms of efficiency and
scalability, even the boldest proposals developed in networking
research. Many open questions remain, but we believe the
applicability of compact routing techniques to Internet interdomain
routing is a research area whose potential payoff for the future of
networking is too high to ignore.

\end{abstract}

\section{Introduction}

The relentless growth of the Internet has brought with it
inevitable, and inextricable, economic and social dependence
on this infrastructure. In reality, the dependence is mutual:
without a robust source of economic support, the Internet will be
unable to evolve toward the potential it quite imaginably holds.
We believe that we will soon be faced with a critical
architectural inflection point in the Internet.  Most experts
agree\footnote{See~\cite{NanogScalability,huston01-03,huston01-02,griffin-wired,routing-requirements}
for few examples.}
that the existing data network architecture is severely stressed
and reaching its capability limits. The evolutionary dynamics of
several critical components of the infrastructure suggest that the
existing system will not scale properly to accommodate even
another decade of growth. While ad hoc patches, some of them
quite clever, have offered temporary relief, both the research and
operational communities concur that a fundamental top-to-bottom
reexamination of the routing architecture is requisite to a
lasting solution.

In this paper we set to work on such reexamination. We clarify up front
that we adhere to standard scientific reductionism
in this work: given a complex system---existing interdomain routing---that
we want to fix, we seek to decompose it into constituents that
we can tackle one-by-one, since each constituent problem
has a chance of admitting formalization and rigorous solution.
We can then construct the final
solution---a future Internet routing architecture---from the solutions
to subproblems. The constituent problem we will
examine in this paper is routing scalability, though it is
well-established that scalability is only one of many
problems of the current Internet routing architecture.\footnote{Other
problems include security, isolation,
configuration control, etc. See~\cite{routing-requirements} for a long
list of future routing architecture requirements.}

More specifically, we analyze the fundamental causes of the global
interdomain routing scalability problems, and how they imply the need for
dramatically more efficient routing algorithms with rigorously
proven worst-case performance guarantees. Fortunately, such algorithms
exist---they
are known collectively as {\em compact routing}. But before we
review these techniques (Section~\ref{sect:compact_routing}),
we first reduce Internet scalability problems to compact
routing formalism. In Section~\ref{sect:scalability} we will peel
off the surface layers of the complexity of the current interdomain
routing system to reveal the roots of its scalability inefficiencies.
One of the serious practical concerns with compact routing is that
it does not guarantee the use of shortest paths: it ``stretches'' them.
We discuss the stretch of a routing
algorithm in Section~\ref{sect:stretch}, and contrast it
with more familiar forms
of path-length inflation in today's interdomain routing. This discussion
naturally leads to our explanation of why {\em hierarchical routing\/} will
not scale if deployed in the Internet (Section~\ref{sect:noscale}).
After we survey the history of compact routing in
Section~\ref{sect:compact_routing},
we analyze the key aspects of its applicability to Internet
interdomain routing in Section~\ref{sect:applicability}.  Such analysis
yields an outline of future work in this area. We conclude
in Section~\ref{sect:conlcusion} with a vivifying summary.

\section{Routing scalability}
\label{sect:scalability}

Poor scaling of a routing protocol expresses itself in terms
of rapid (linear) rates of growth of the routing table~(RT) size
as well as convergence parameters (convergence time, churn,
instabilities, etc.).  In this paper we concentrate on the
subproblem associated with the former component, but we also
discuss the latter in Section~\ref{sect:applicability}.

The immediate causes of Internet RT growth have been extensively
studied and analyzed~\cite{huston01-03,huston01-02,BroNeCla02,NaGoVa03}.
Several studies, proposals, and even new routing architectures seek to
directly address these causes. Prominent efforts in this area
include~\cite{beyond-bgp,atoms,multi6,islay,nimrod} and most
recently~\cite{GuKoKi04,SuKaEe05}.
However, we argue that even if
we could manage to deploy any of the proposed approaches, it would
represent only a short-term patch, because they address
symptoms rather than the root of the problem. It is empirically
clear that the Internet needs more than a short-term fix.

As an example, we consider the idea of
{\em routing on AS numbers}, where AS numbers replace IP prefixes
in the role of interdomain routing addresses.
This idea is now a common part of many
proposals~\cite{atoms,islay,nimrod,GuKoKi04,SuKaEe05}.
At first glance routing on AS numbers looks like a final
solution to the RT size problem.  Indeed, such an approach
could immediately reduce the RT size by an order of magnitude,
since there are an order of $10^5$ IP prefixes and $10^4$ ASs
in the global RT today.  However, even if a protocol implementing
this idea were deployed, there is no reason to believe it would
change the empirically observed trend\footnote{In the existing Internet,
the total number of ASs grows faster than the total number of IP
prefixes~\cite{huston-bgp-site,BroNeCla02}.}
that an increasing number of small (stub) networks want to participate
in interdomain routing, e.g., for traffic engineering reasons,
and to do so they require their own interdomain routing addresses
(AS numbers). Such small networks utilize tiny portions of
the public IP address space,
which implies that inevitably the total number of ASs will
continue to grow at roughly the same rate as the total number
of IP prefixes, and we will soon face, albeit a few years later,
essentially the same problem we have today.

\begin{figure}[t]
    \centering
    \includegraphics[width=3in]{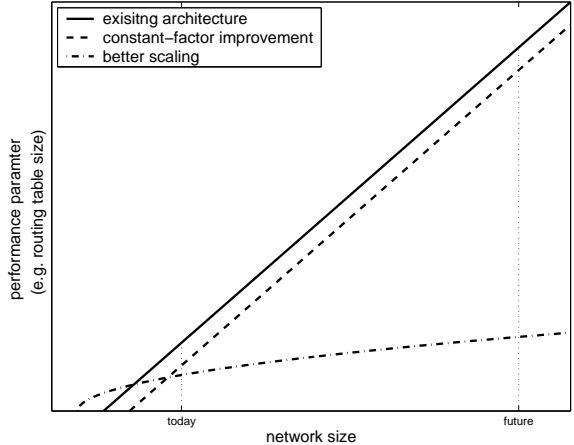}
    \caption{\footnotesize {\bf Constant-factor improvement vs.~better scaling.}
    Scalability means effective scaling, not a one-time reduction.
    Although the difference in impact between alternatives to
    the existing architecture may be small today, in the long run
    a constant factor reduction is negligible compared to the
    performance gain from a genuinely scalable solution.
    Any solution that scales in essentially the same way as
    the existing architecture is only postponing the problem.
    \label{fig:scaling}}
 \end{figure}

We learn two important lessons from this example. The first is of
general significance: {\bf when considering scalability aspects of a
new routing architecture, what matters is not any one-time improvement
it produces, but how well it scales}.  A routing protocol that yields
a constant reduction in RT size but the same rate of growth does not
actually solve the RT size problem, it simply postpones it.
Figure~\ref{fig:scaling} illustrates this consideration: as
network size grows, the difference between the current architecture
and an architecture providing a constant-factor improvement becomes
negligible.  No less important is the {\em mathematical provability\/}
of the protocol's scaling behavior. In the case of RT size
such provability implies rigorously derived worst-case (upper) bounds
of the RT size as a function of the network size.
The fact that we face interdomain routing scalability problems
today is at least in part due to the lack of any rigorous performance
guarantees embedded in the BGP design. We should not repeat the same
mistake in the future.

The second lesson is more specific: organization (AS) boundaries do
offer a natural level of aggregation and abstraction of routing
information. It would be thus unexpected to neglect such a possibility
for RT size reduction. Indeed, we adopt the idea of routing on AS
numbers, and the natural level of network topology abstraction we
consider in our work is thus the {\em Internet AS-level topology graph}.

Furthermore, faithful to our reductionistic spirit, we temporarily
avoid discussing routing policies. T.~Griffin {\it et al.}'s independent
scope of work focuses specifically on policy-related problems.
In addition, one design goal of their latest abstract routing
protocol model~\cite{GriSo05} is explicit modularity with respect to
the routing {\em algorithm\/} and {\em policy\/} components of the
protocol.\footnote{Note
the difference between terms {\em routing protocol\/} and
{\em routing algorithm}.
The algorithm is a part of a protocol. In BGP case, for example, the protocol
is BGP itself, while algorithm is path-vector.}
Thus we believe that our work on the algorithmic component
can proceed in parallel and virtually independently of the
vitally important policy-related research. Satisfactory
results in both domains would powerfully position the
research community to offer a complete solution.

To summarize this section, we assert that in order to find sustainable
solutions to the global Internet routing problem, we need to
investigate the scalability issues at their most fundamental level.
For interdomain routing the framing question appears to be the
performance guarantees of routing algorithms operating on graphs with
topologies similar to the Internet AS-level topology.  In particular,
we are interested in rigorous upper bounds of the RT size.
It turns out that compact routing is
exactly what we are looking for:
a research area focused on construction of efficient routing algorithms with
proven worst-case performance guarantees. However, before we describe compact
routing, we need to discuss stretch.

\section{Routing stretch}
\label{sect:stretch}

The {\em stretch} of a routing algorithm
is defined as a worst-case multiplicative path-length increase factor.
More specifically, for every pair of nodes in all graphs
in the set of graphs the algorithm can operate on,
we find the ratio of the length of the path produced by the
algorithm to the length of the shortest path between the same pair
of nodes. The maximum of this ratio among all node pairs in all
the graphs
is the algorithm's stretch.
The average stretch is the average of this ratio on a subset of graphs.

We emphasize that {\bf stretch has nothing in common with known
forms of path inflation} in contemporary Internet
routing. Path inflation is today's Internet is {\em not\/} due to the non-trivial
stretch of an underlying routing algorithm, but rather due to
intra- and inter-domain routing policies, various incongruities
across multiple levels of network topology abstraction
(geographical, router-, AS-level), etc.~\cite{SprMaAn03}

In fact, we are not aware of any currently used routing protocol
that would implement a routing algorithm with stretch greater
than~1 (stretch$>$1): link-state (LS) in OSPF or ISIS,
distance-vector (DV) in RIP, LS/DV hybrid in EIGRP, and path-vector in
BGP are all forms of
{\em trivial}\footnote{The word {\em trivial} is not a judgment
but a reference to {\em trivial shortest-path
routing\/} in Section~\ref{sect:compact_routing}.}
shortest-path routing (stretch-1).  Consider two examples:
BGP and OSPF.   Non-trivial policy configurations generally
prevent BGP from routing along the actual shortest path,
but without policies the BGP route selection process
would always select the shortest paths in the AS-level graph.
Similarly, non-trivial area configurations prevent OSPF
from routing along the shortest paths, but routing inside
an area is always along the shortest path in the weighted router-level graph.

The above examples seem straightforward but we identify them
to eliminate any possible confusion between stretch of a
routing {\em algorithm\/} and inflation of paths produced by a routing
{\em protocol}.  Clarifying the difference allows us to formulate
both necessary and sufficient requirements for routing stretch.
The {\em sufficient routing stretch requirement} is that
{\bf the routing algorithm underlying
any realistic interdomain routing protocol must be of stretch-1.}
Our reasoning behind this requirement is as follows.
Consider a {\em generic}\footnote{A {\em generic} routing algorithm
works correctly on all graphs.} stretch$>$1 routing algorithm and
apply it to a complete graph, which has all nodes directly
connected to all other nodes so that all shortest paths are of length~1.
If the routing algorithm does not guarantee to find all shortest paths,
it would imply that some nodes in a complete graph would
not know about their own directly connected neighbors.

Of course, the Internet interdomain topology is not a full mesh,
and it might turn out that the same routing algorithm that fails
to be stretch-1 on a full mesh does in fact find all shortest paths
to each node's neighbors in realistic Internet-like topologies.
We thus might also pursue the explicit task of finding
a non-generic routing algorithm that is applicable only to Internet-like
topologies. We do not require this algorithm be stretch-1, but it
must be of stretch-1 on paths of length~1 leading to nodes' neighbors.
In other words, the {\em necessary routing stretch requirement}
is that {\bf the routing algorithm underlying any realistic
interdomain routing protocol must be of stretch-1 on paths
of length~1 in realistic Internet-like topologies.}

These requirements are purely practical: if a link exists between
a pair of ASs, they must be able to route traffic over this
link. If our sophisticated routing algorithm removes routing
information about this link from the ASs' RTs, then the ASs'
network operators will have to manually reinsert it, which counters
the scalability objective of reducing the RT size. We will refer to
this administrative re-adding of information about shortest
paths to nodes' neighbors as {\em neighbor reinsertion}.

The absence of a neighbor of a node in that node's RT appears
to contradict common sense, but one can verify that it
occurs with the algorithms we discuss in
Section~\ref{sect:compact_routing}. We emphasize
that we have not had to deal with this effect in practice yet
because the routing algorithms in operation today are all of stretch-1.
Indeed, this effect reflects the inevitable trade-off between
RT size and stretch.  Intuitively, we can explain this trade-off
as follows: shortest path routing implies complete knowledge of the
network topology, complete knowledge suggests a lot of information,
and a lot of information suggests large RTs. In order to shrink RTs,
we must prepare to lose topological information, and such loss
necessarily increases path lengths.  In addition, the meshier the
network, the more information about node neighbors there is to lose.
The full mesh is an extreme example of this dilemma.

This fundamental and unavoidable trade-off between stretch of a
routing algorithm and sizes of RTs it generates is at the center of
compact routing research going on today, but the first formalization of
this trade-off was introduced many years ago in the context of
hierarchical routing.

\section{Why hierarchical routing will not work}
\label{sect:noscale}

The pioneering paper on routing scalability was
Kleinrock and Kamoun's~\cite{KK77}, which introduced the
idea of hierarchical routing. The key idea is to group (or aggregate)
nearby nodes into areas, areas into super-areas, and so on.
Such grouping is a form of the {\em network partitioning\/} problem.
Hierarchical aggregation and addressing offer
substantial RT size reduction by abstracting out
unnecessary topological details about
remote portions of the network: nodes in one \mbox{(super-)area} need to keep
only one RT entry for all nodes in another \mbox{(super-)area.}
To our knowledge, all proposals for future Internet routing
architectures trying to address the RT size problem are based,
explicitly or implicitly, on this concept of hierarchical routing. In
fact, no other kind of RT size reduction technique has
been considered in the networking literature.\footnote{Recall
that we have already adopted routing on AS numbers,
so that we consider here the RT size reduction {\em below\/}
the total number of ASs in the interdomain topology graph.}
Unfortunately, according to the best available data,
hierarchical routing is simply not a good candidate for
interdomain routing.

Indeed, in the same paper~\cite{KK77}, Kleinrock and Kamoun were the
first to analyze the stretch/RT size trade-off. They showed
that the
routing stretch produced by the hierarchical approach
is satisfactory only for
topologies with average shortest path hop-length (distance)~$\bar{d}(n)$
that grows quickly with network size~$n$, which means
it works well only for graphs where long paths prevail.
In fact, they assumed \mbox{$\bar{d}(n) \sim n^\nu$} with~$\nu>0$.
Such graphs have many remote nodes, i.e., nodes at long hop distances
from each other, and one can efficiently, without substantial
multiplicative path length increase, aggregate details about topology
around remote nodes, simply because those remote nodes are abundant.

Unfortunately, according to the best available data,
the Internet topology does not meet those conditions.
The Internet AS-level topology is a {\em scale-free\/}
network with characteristics described, for example,
in~\cite{MaKrFoHuDiKcVa05-tr}.
To avoid terminology disputes,  by {\em scale-free\/}
we simply mean networks with fat-tail, e.g., power-law,
node degree distributions. In theory, the average
distance in such topologies grows with network size
much more slowly than~\cite{KK77} assumed:
it is at most \mbox{$\bar{d}(n) \sim \log n$}~\cite{ChLu03}. In practice,
the average hop distance between ASs in the Internet stays virtually
constant or even decreases due to increasing inter-AS
connectivity~\cite{huston-bgp-site,BroNeCla02}. According
to multiple data sources, the average AS-hop distance
in the current Internet is between~3.1 and~3.7,
with more than~80\% of AS pairs~\mbox{2-4} hops away from each
other~\cite{MaKrFoHuDiKcVa05-tr}.
We call a network {\em small-world\/} if most of its nodes are at small
distances from each other.
Thus the Internet AS-level topology is small-world:
{\em it has essentially no remote nodes}, which is extremely
bad news since the effectiveness of network partitioning
depends on their abundance.
In other words,
{\bf the unsettling but plainly observed reality is that
one cannot efficiently apply hierarchical routing to small-world
Internet-like topologies}.

The previous arguments might sound too general since
we could still apply some generic hierarchical routing algorithm
to the Internet. However, such application cannot be efficient and
we must prepare to see high stretch.
And indeed, simple analytical estimates~\cite{KrFaYa04}
show that applying hierarchical routing to an Internet AS-level
topology incurs a $\sim$15-time path length increase, which,
although alarming enough by itself, would also lead to a substantial
RT size surge caused by neighbor reinsertion.
If we accept our argument in Section~\ref{sect:stretch}
that any truly scalable Internet routing algorithm must have
stretch as close as possible to~1, we must also accept the
fact that hierarchical routing will not meet our needs.
We conclude that all previous Internet interdomain routing
proposals heavily based on hierarchical routing
ideas\footnote{Classic examples of such proposals are~\cite{nimrod,islay}.
Note that recent work in~\cite{SuKaEe05} advocates hierarchical
routing, but the authors do not try to use it to reduce RT size.
They do reduce RT size, but only by routing on AS numbers.}
are not realistic and, if deployed, would not genuinely scale.

\section{History of compact routing}
\label{sect:compact_routing}

Fortunately, Kleinrock and Kamoun's results~\cite{KK77} are the
first but not the last results of research on scalable routing.
The history of this research is not linear; we outline it in
this section.

We have not mentioned yet that~\cite{KK77} also suffered from
a problem more serious than high stretch on small-world
graphs. It did not provide any algorithms to actually
construct, for a given topology, a network partitioning
upon which hierarchical routing can operate.
Therefore, in works that followed~\cite{KK77}, first
L.~Kleinrock himself, then R.~Perlman, and then in the late
1980s and in the 1990s B.~Awerbuch, D.~Peleg and others,
all spent much time and effort trying to salvage hierarchical
routing and to find efficient network partitioning algorithms.
It did not become clear that hierarchical routing does not have
any future until~1999 when L.~Cowen~\cite{cowen99}
delivered her brilliant compact routing scheme.
Her scheme had four
attractive features: 1)~it was non-hierarchical, so required
no network partitioning; 2)~it was generic and so worked for
any topology; 3)~it had a fixed stretch of 3 for any topology of
any size;
and 4)~it had a sub-linear RT size upper bound of
$\tilde{O}(n^{2/3})$. Cowen's algorithm was also beautifully
simple compared to numerous hierarchical routing algorithms
accumulated by that time.

In retrospect, we can say that compact routing, which is
the state-of-the-art in efficient routing algorithm research today,
built on numerous findings from hierarchical routing,
which had been the state-of-the-art for decades.
Today, compact routing is a research
area in the theory of distributed computation, a part of theoretical
computer science. There is no strict definition of a compact routing
algorithm, or {\em scheme}, but it usually means an algorithm with
a sublinear RT size
upper bound---the RT size is also called {\em local memory space},
or simply {\em space} in compact routing terminology---and
with stretch bounded by a constant, in contrast to many
hierarchical routing algorithms for which stretch grows to infinity
with network size.\footnote{One can verify that on an $n$-node full mesh,
for example, \cite{KK77}~produces stretch growing to infinity as~$\Theta(\log n)$.}

Since 1999 progress has been much more rapid than in the preceding~22
years. In 2001, M.~Thorup
and U.~Zwick~\cite{ThoZwi01b}~(TZ) improved on Cowen's RT size (space)
upper bound, bringing it to~$\tilde{O}(n^{1/2})$ while maintaining stretch of~3.
This result was particularly significant because the authors had previously
proven~\cite{ThoZwi01a} that any routing with stretch
below~5 cannot guarantee space smaller than~$\Omega(n^{1/2})$.
Therefore the TZ scheme was the first stretch-3 scheme whose
memory space upper and lower bounds were nearly the same
(up to a poly-logarithmic factor implied by~\mbox{` $\tilde{}$ '} in
the $\tilde{O}$ notation). In other words, the TZ scheme was
the first generic nearly {\em optimal} routing scheme of stretch~3.

The main problem with both the Cowen and TZ schemes is that they
are name-dependent, meaning that topological information is embedded
in node addresses which thus cannot be arbitrary.
Such schemes cannot be used in dynamic topologies, since any
topology change potentially requires renaming many nodes.
In 2003, M.~Arias {\em et al.}~\cite{ArCoLaRaTa03}, delivered
a {\em name-independent}, i.e., name/addresses of nodes can be
arbitrary, routing scheme with a memory space upper bound
of~$\tilde{O}(n^{1/2})$ and stretch of~5.
In 2004, I. Abraham {\em et al.}~\cite{AbGaMaNiTho04} improved
on~\cite{ArCoLaRaTa03} by decreasing stretch to~3, which makes
it the first generic nearly optimal {\em name-independent\/} stretch-3
routing scheme.

Why do we mention stretch-3 so often?
For \mbox{stretch-1} (shortest path) routing,
one can construct RTs at each node
by storing, for every destination node, the ID of the
outgoing port on the shortest path to the destination.
The number of destinations is \mbox{$n-1$}, the maximum number of ports
a node can have (equal to maximum possible node degree)
is also \mbox{$n-1$}, and thus a maximum $O(n \log n)$
bits of memory is required for an RT. This
construction is called {\em trivial shortest path routing}.
However, in 1996, C.~Gavoille and S.~P\'{e}renn\`{e}s~\cite{GaPe96}
showed that the {\em lower\/} bound of generic stretch-1
routing is also $\Omega(n \log n)$, i.e., there is no shortest
path routing scheme that guarantees local memory space smaller than
$\Omega(n \log n)$ at all nodes of all topologies.
In other words, shortest path routing is incompressible.
In order to decrease the maximum RT size, we
must allow maximum stretch to increase. Finally,
in 1997, C.~Gavoille and M.~Genegler~\cite{GaGe97} showed that
for any stretch strictly below~3, the local space
lower bound is nearly the same as for stretch-1 routing,~$\Omega(n)$,
meaning that {\em no\/} generic stretch$<$3 routing scheme can
guarantee sublinear RT sizes: the minimum value of maximum stretch
allowing sublinear RT sizes is~3.

\section{Compact routing applicability}
\label{sect:applicability}

The end of the last section sounds alarming: it conflicts with
considerations in Section~\ref{sect:stretch}, where we
discuss why stretch should be close to~1 for practical
applicability. Do shortest path routing incompressibility and
impossibility of sublinear-space \mbox{stretch$<$3} routing mean
that scalable interdomain routing with sublinear RT sizes is not
achievable at all?

To answer this question, recall from Section~\ref{sect:stretch}
that stretch refers to the {\em maximum\/} path length increase among all
paths in {\em all the graphs\/} on which a routing algorithm operates,
meaning that there exists some worst-case graph\footnote{See~\cite{GaGe97}
for the explicit construction of this graph.} on which this
maximum is achieved. Recall also that all schemes mentioned
in Section~\ref{sect:compact_routing} are generic; they can
work on {\em any\/} graph, meaning that the worst-case graph does not have
to be Internet-like. We can consequently expect the {\em average
stretch\/} of these schemes on the class of Internet-like
graphs to be lower than~3. But how close can it be to~1? At first
thought it seems unlikely to be close to~1 at all, since we saw
in Section~\ref{sect:noscale}
that the stretch of hierarchical routing
tends to be high on small-world graphs, and there is no reason to
believe that compact routing would be drastically better. However,
in 2004, \cite{KrFaYa04}~showed that the
average performance of the first optimal stretch-3 scheme~(TZ)~\cite{ThoZwi01b}
on Internet-like topologies is spectacularly better than its worst case:
while its upper bounds are around~2200 for RT size and~3 for stretch,
{\bf the average RT size of the TZ scheme on the real Internet topology
was found to be around~50 entries (for the whole Internet!)
and the average stretch was only~1.1!}\footnote{We remind that while
these numbers are overwhelming, they correspond to a one-time improvement.
In the long-term, what matters is the mathematically
proven scaling behavior of the scheme (cf.~Section~\ref{sect:scalability}).}
These numbers are striking evidence that, on the same topology and
under the same assumptions,
compact routing dramatically outperforms both hierarchical
routing~\cite{KK77}, with its stretch of~15 and RTs of unknown sizes,
and simple routing on AS numbers (e.g.,~\cite{SuKaEe05}),
with RT sizes of the order of~$10^4$.

An interesting question is why compact routing turns out to be
so spectacularly efficient on Internet-like topologies. There is no
rigorous answer to this question yet, but intuitively
we can think of a scale-free network as a dense interconnection
of star graphs, and one can verify that the stretch of many
compact routing schemes, e.g., Cowen, TZ,
on stars is~1. In addition, the scalability of the most
efficient routing on trees---and stars are trees---is
dazzling: RT sizes and RT lookup times are nearly constant
(they do not grow with network size), and stretch is~1~\cite{ThoZwi01b}!
Of course, the Internet topology is not at
all tree-like, since strong
clustering~\cite{MaKrFoHuDiKcVa05-tr} renders its treewidth high.
But surprisingly, \cite{fraigniaud05}~finds that strongly clustered
networks and networks with low
treewidth both possess the set of properties allowing for
remarkable performance of an important class of routing strategies.

While the preliminary findings in~\cite{KrFaYa04} clearly demonstrate the
potential power and superiority of compact routing to everything
previously considered in networking in pursuit of routing
scalability, many open questions remain. In the remainder
of this section we catalog the four open problems we consider
most important.

The average stretch was found to be close to~1 in~\cite{KrFaYa04},
but not~1. In fact, the average stretch does not even approach~1
in the limit of large networks.  One can verify that if the average
stretch does not approach~1 on scale-free graphs at least as fast as
$1+\tilde{\Theta}(n^{-1/2})$, then the average number of non-shortest
paths per node is such that the neighbor reinsertion from Section~\ref{sect:stretch}
will break the theoretical RT size upper bound of~$\tilde{O}(n^{1/2})$.
If the average stretch does not approach~1 at all, as in~\cite{KrFaYa04},
then neighbor reinsertion forces the RT size to
go from~$\tilde{O}(n^{1/2})$
back to its trivial linear scaling~$\tilde{O}(n)$.  Therefore, {\bf {\em the
stretch scaling problem\/}
is to find a compact routing scheme with satisfactory
scaling of stretch}, at least for length-1 paths.
Such a scheme likely exists given our intuition behind why the
stretch of the TZ scheme
on the Internet is so low in the first place: larger scale-free networks
are increasingly ``star-like''.

A stronger formulation of the same problem is to find {\em shortest path\/}
routing with sublinear RT sizes on scale-free graphs. Sublinear-space
\mbox{stretch-1} routing applicable to all graphs cannot exist
(cf.~Section~\ref{sect:compact_routing}), but if we narrow
the class of graphs to scale-free graphs then there are
no results so far that prove the non-existence of such routing.
In addition,
it may turn out that we can significantly decrease RT size upper bounds,
a likely possibility given how far below its upper bound
the RT size of the TZ scheme is in~\cite{KrFaYa04}. In more general terms,
{\bf {\em the scale-free routing problem\/}
is to utilize peculiarities of scale-free networks to obtain
non-generic algorithms optimized specifically for Internet-like topologies}.

{\bf The most vital open problem,
{\em the dynamic routing problem}, is to construct
truly dynamic schemes}. All currently existing schemes
are {\em static}. A {\em dynamic\/} scheme would have rigorous
upper bounds for its performance parameters
in dynamic networks, where nodes
and links can fail, be added or removed. Dynamic performance parameters
include the number of control messages the algorithm generates to converge after
a topology change, sizes of those messages, number of affected nodes, etc.
We refer to these parameters collectively as {\em convergence costs}.
Note that all currently deployed routing algorithms are
essentially static algorithms applied to dynamic environments,
an approach we know can be costly, e.g., the convergence costs
of BGP are infinite given that it can oscillate persistently and not
converge at all~\cite{GriWil99,GriWil02}.
The convergence costs of BGP's routing algorithm
(path-vector) are also astronomically high,~$O(n!)$~\cite{LaAhBo01}.
We need a solution with satisfactory bounds of convergence costs.

Unfortunately, this problem is fundamentally hard, as illustrated by
the notable but pessimistic result~\cite{AfGaRi89} which says that
even with unbounded RT sizes, there is no generic dynamic stretch-$s$
routing scheme that guarantees less than~$\Omega(n/s)$ routing
messages.  Simply put, there is no generic routing scheme that
scales sublinearly in the number of control packets per topology
change.\footnote{Although linear scaling in practically unacceptable,
it is still much better than the $O(n!)$ scaling of the path-vector
algorithm.} As in the static case we hope that narrowing the
problem from the class of all graphs to the class of Internet-like
graphs will yield more auspicious scaling.

An alternative shortcut is to disregard the fact that all existing compact
routing schemes are static and to test them in dynamic networks anyway.
The intuition behind why these schemes might still work well even
in the dynamic environment lies in their ultra-optimal static performance.
Since the existing {\em name-dependent\/} schemes are not appropriate
candidates for such experimentation for
reasons we mentioned in Section~\ref{sect:compact_routing}, {\bf
{\em the name-independent routing problem\/} is to analyze the
performance of static name-independent routing schemes on dynamic
Internet-like topologies}. Work
in this direction has already begun: in~\cite{AbBaBi04} the authors
tested the first optimal \mbox{stretch-3} name-independent routing
scheme~\cite{AbGaMaNiTho04} which they implemented and deployed
in an overlay on PlanetLab in October~2004.

\section{Conclusion}
\label{sect:conlcusion}

The ultimate goal of our work is to build an incrementally
deployable interdomain routing {\em protocol\/} that will encompass
not only a new routing {\em algorithm\/} with satisfactory rigorous
bounds for routing table size, stretch, and convergence costs, but
also routing {\em policies} with appropriate performance guarantees.
We believe it is too early to discuss protocol details since we do
not yet even have constituents with acceptable characteristics, much
less understand their idiosyncrasies.

On the positive side, this quest has already begun.  Current
circumstances in the fields of theoretical computer science (TCS) and
networking have created a rare opportunity to pursue a fundamental
breakthrough in the search for truly scalable routing. Recent
developments in TCS yield shockingly promising results: alternative
{\em compact routing} algorithms offer dramatically superior
scalability characteristics than those being used or even considered
in networking.  We recognize the need for a radical change in our
approach to routing research in order to build on these contemporary
results from TCS. The first task is to determine how applicable
these results are to real-world networks.

Construction of practical, efficient, and scalable routing for the
next-generation Internet is a key motivator of our work, but we
recognize that we are only at the beginning of an extensive analysis
of our most fundamental assumptions about routing in data networks.
One such assumption is the adequacy of representing a real dynamic
network using an intrinsically static construct, a graph. This
assumption underlies, either explicitly or implicitly, all forms of
routing we have discussed in this paper. If we discover in the
future that scalable dynamic routing cannot even exist in principle
within the static graph-theoretic framework constraints, then some
unknown but presumably paradigm-shifting approaches to data network
routing will become inevitable.  Regardless of the outcome, what
matters today is that we have a unique opportunity to explore a
radical new path, which initially looks more promising than any
previously considered approach to scalable routing.

\section{Acknowledgements}

This work is supported by NSF CNS-0434996.

\bibliographystyle{unsrt}
\bibliography{bib}

\end{document}